# High-Q Dual-Band Graphene Absorbers by Selective Excitation of Graphene Plasmon Polaritons: Circuit Model Analysis

## Saeedeh Barzegar-Parizi[1*], Amir Ebrahimi[2], Kamran Ghorbani[3]

[1]Department of Electrical Engineering, Sirjan University of Technology, Sirjan, Iran

[2,3] School of Engineering, RMIT University, Melbourne, Australia.

Email:barzegarparizi@sirjantech.ac.ir

## Abstract

This article presents the design of two dual-band graphene-based absorbers for terahertz frequencies. The absorbers are composed of two-dimensional (2D)arrays of ribbons and disks printed on a ground plane backed dielectric spacer. The design is based on the excitation of a specific plasmon polariton of the graphene patterned array at each resonance band. An analytical circuit model is used to derive closed-form relations for the geometrical parameters of the absorber and graphene parameters. The graphene patterned array appears as a surface admittance made of an infinite parallel array of series RLC branches. Each branch is equivalent to a graphene plasmon polariton (GPP) providing a distinct resonance mode. The design procedure is based on selectively exciting the first two GPPs. This means that two of the parallel RLC branches are selectively used in the circuit model analysis. The results obtained using the analytical solution are compared with the full-wave simulations in HFSS. The agreement between the results validates the developed design method.

## 1. Introduction

Surface plasmon polaritons (SPPs), the eigenmodes of an interface between a dielectric and metal with strong interaction between light and free electrons, have attracted considerable attention in the past decades [1-3]. SPPs allow overcoming the diffraction limit to confine light into deep-subwavelength volumes with huge field enhancements [2,3]. Noble metal-based plasmonics with fascinating properties found promising applications in switches [4,5], demultiplexers [6,7], and photovoltaics [8,9] in visible and near-infrared regimes. However, the noble metals-based plasmonic devices are constrained due to some limitations such as poor confinement, difficulty in controlling their permittivity functions, and high material losses at teraherhz (THz) frequencies.

The discovery of graphene emerged new opportunities in optoelectronics and photonics in mid-infrared and terahertz regions [10,11]. Graphene, a one-atom-thick sheet of carbon, offers a novel design platform due to its attractive electrical and optical properties such as high thermal conductivity [11], Gate-variable optical conductivity [12], controllable plasmonic properties [13], and high-speed operation [14]. Like noble metals, graphene is a promising candidate for plasmonic devices. However, graphene plasmon polaritons (GPPs) offer strong confinement and relatively longer propagation distance [15]. In addition, the plasmonic properties of graphene can be controlled electrically or chemically [16-18]. These features make it a promising candidate for many applications such as optical modulators [19-21], sensors [22-26], and absorbers [27-39].

Extensive studies are performed so far on the graphene-based absorbers for broadband and multiband applications [27-40]. Absorbers have applications in solar energy harvesting [41,42], refractive index sensors [43], microbolometers [44], thermal imaging [45], thermal IR emitters [46]. Here, two-dimensional patterned graphene ribbon and disk arrays are used to design dual-band absorbers. In our previous dual-band absorber designs [39,40], fundamental GPPs were excited at both of the absorption bands. However, the current work discusses the design of dual-band absorbers by selectively exciting a specific GPP at each frequency band. This is performed by developing an analytical procedure that returns the required geometrical parameters and graphene properties such as chemical potential and relaxation time for excitation of a specific GPP. The proposed absorbers show higher absorption quality factor (Q) in comparison with state-of-art graphene-based dual-band absorbers. Furthermore, the new design offers more absorption stability with respect to the oblique incidence angles and provides lower out-of-band absorption values between the adjacent channels.

A circuit model-based procedure is developed for the design of the proposed absorbers providing an analytical solution for the initial design of the unit cell geometrical parameters. This obviates the requirement for time-consuming full-wave simulations for optimization of the absorber response. We demonstrate that the resonance frequencies can be controlled by varying the geometrical parameters or chemical potential of the graphene layer.

The rest of the paper is organized as follows: Section 2 presents an analytical circuit-based design of a dual-band absorber based on an array of graphene ribbons. Closed-form equations are derived for the absorber parameters by exciting two GPPs of the graphene ribbons array. The absorber design examples are presented in this section. The ribbon array-based absorber is polarization sensitive. Thus, Section 3 presents a polarization insensitive dual-band absorber based on an array of graphene disks followed by discussions and comparisons with the state-of-art designs in the literature. Finally, Section 4 presents the main conclusions.

## 2. Dual-band absorber based on an array of graphene ribbons

Electromagnetic absorbers are usually implemented in a three-layer configuration made of a periodic array deposited on a dielectric film terminated by a metallic ground plane. The top periodic array contains one or two-dimensional subwavelength patterned elements. The absorption is calculated as $A(\omega) = 1-R(\omega)-T(\omega)$, where $T(\omega) = |S_{21}|^2$ and $R(\omega) = |S_{11}|^2$ are the transmission and reflection coefficients, respectively. The ground plane prevents transmission of the electromagnetic waves through the structure. Hence, the absorption is calculated as $A(\omega) = 1-R(\omega)$.

In [47], [48], the GPPs were studied in the arrays of graphene ribbons and disks, where the corresponding eigenvalues and eigenfunctions are analytically extracted. Here, we design the geometrical parameters of the absorber and the properties of graphene to selectively excite two GPPs to achieve a dual-band absorber with nearly perfect absorption.

Consider an absorber configuration in figure 1. The absorber is made of three layers: a top layer of graphene ribbon array, a middle dielectric spacer, and a metallic layer acting as a ground plane. The width of the graphene ribbons is $w$. The ribbons are periodically arranged along $x$-direction with a period of $L$, whereas they are infinitely long in $y$-direction. The dielectric spacer is with thickness of $h$ and refractive index of $n_s$. The surface conductivity of graphene is given by [48]:

$$\sigma_g = \frac{2e^2 k_B T}{\pi \hbar^2} \frac{j}{-\omega + j\tau^{-1}} \ln\left[2\cosh\left(E_F / 2k_B T\right)\right]$$
$$- \frac{je^2}{4\pi\hbar} \ln\left[\frac{2E_F - \hbar(\omega - j\tau^{-1})}{2E_F + \hbar(\omega - j\tau^{-1})}\right],$$
(1)

where $e$ is the electron charge, $E_F$ denotes the Fermi level, $\hbar$ refers to the reduced Plank constant, $k_B$ represents the Boltzmann constant, $\omega = 2\pi f$ is the angular frequency, $T = 300$ K is the room temperature, and $\tau$ is the relaxation time.

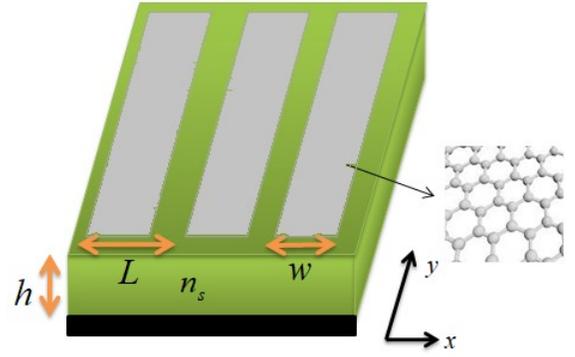

**Figure 1**. Structure of an absorber based on an array of graphene ribbons patterned on a dielectric spacer with thickness of $h$ and refractive index of $n_s$. The dielectric spacer is terminated with a metallic film acting as a ground plane. The graphene layer is shown in grey, the dielectric space is shown in green, and the ground plane is shown in black color.

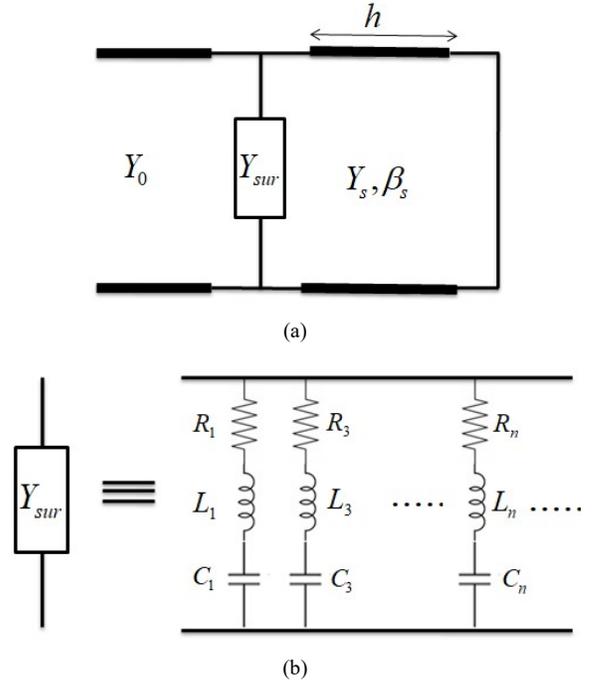

**Figure 2**. (a) Equivalent circuit model of the proposed absorber. (b) The parallel branches composed of RLC series tanks corresponding to the equivalent surface admittance of the ribbons array.

The surface conductivity of graphene contains intraband conductivity represented by the first term in (1) and interband conductivity modelled by the second term in (1). Suppose that the absorber is illuminated by a TM polarized plane wave. The equivalent circuit model of the absorber is presented in figure 2(a) for this case. In the circuit model, the ribbon array is modelled by a surface admittance as [47]:

$$Y_{sur} = \sum_{n=1(odd)}^{\infty} \frac{S_n^2}{D} (\sigma_g^{-1} + \frac{q_n}{2j\omega\varepsilon_{eff}})^{-1},$$
(2)

where $q_n$ is the $n$'th eigenvalue and

$$S_n = \int_{-w/2}^{w/2} \psi_n(x)\,dx.$$
(3)

with $\psi_n(x)$ being the $n$'th normalized eigenfunction corresponding to $n$'th resonance modes of the ribbons array.

The eigenfunctions and eigenvalues are calculated in [47]. In (2), $\varepsilon_{\text{eff}} = \varepsilon_0(1 + n_s^2)/2$. The even modes do not appear in (2) since $S_n$ values are zero for the even modes [47]. The surface admittance in (2) can be considered as a combination of parallel branches of series $RLC$ tanks corresponding to each of the GPPs as shown in figure 2(b). Thus,

$$Y_{sur} = \sum_{n=1(odd)}^{\infty} (R_n + jL_n\omega + \frac{1}{jC_n\omega})^{-1}, \qquad (4)$$

where $R_n$, $L_n$ and $C_n$ are defined as:

$$R_n = \frac{L}{S_n^2} \text{Re}\{\sigma_g^{-1}\}, \qquad (5.a)$$

$$L_n = \frac{L}{S_n^2} \frac{\text{Im}\{\sigma_g^{-1}\}}{\omega}, \qquad (5.b)$$

$$C_n = \frac{S_n^2}{L} \frac{2\varepsilon_{\text{eff}}}{q_n}. \qquad (5.c)$$

The aim is to determine the geometrical parameters and graphene properties to excite the first two GPPs. Therefore, $n=1$ and $n=3$ should appear in (2). As explained in [47], the eigenfunctions corresponding to these modes are calculated as:

$$\psi_1(x) = w^{-\frac{1}{2}}\left[1.2\sin(\cos^{-1}(\frac{2x}{w})) - 0.106\sin(3\cos^{-1}(\frac{2x}{w}))\right], \qquad (6.a)$$

$$\psi_3(x) = w^{-\frac{1}{2}}\left[0.308\sin(\cos^{-1}(\frac{2x}{w})) + 1.19\sin(3\cos^{-1}(\frac{2x}{w})) - 0.484\sin(5\cos^{-1}(\frac{2x}{w}))\right], \qquad (6.b)$$

where the corresponding eigenvalues are $q_1 = 0.42\pi/w$ and $q_3 = 2.606\pi/w$ for $w/L = 0.9$ filling factor.

In figure 2(a), $\beta_s = k_0 n_s$ and $Y_s = n_s/\eta_0$ are the propagation constant and the admittance of the transmission line corresponding to the dielectric spacer, where $\eta_0 = 120\pi$ is the free-space impedance. The wave number is $k_0 = \omega/c$ with $c$ being the speed of light in the free space. The metallic ground plane is modelled as a short circuit. So, the input admittance of the circuit is obtained as:

$$Y_{in} = Y_{sur} - jY_s \cot(\beta_s h). \qquad (7)$$

Our purpose is designing a dual-band absorber by exciting a specific GPP at each band. Therefore, the input admittance of the absorber can be written as:

$$Y_{in} = (R_1 + jL_1\omega + \frac{1}{jC_1\omega})^{-1} + (R_3 + jL_3\omega + \frac{1}{jC_3\omega})^{-1} - jY_s\cot(\beta_s h). \qquad (8)$$

At the lower absorption band, the dominate resonance mode of the graphene patterned array is the first GPP and the influence of the higher-order modes is negligible. Therefore, the second term in the right side in (8) can be omitted resulting in:

$$Y_{in}\mid_{\omega=\omega_1} = (R_1 + jL_1\omega_1 + \frac{1}{jC_1\omega_1})^{-1} - jY_s\cot(\beta_s h)\mid_{\omega=\omega_1}, \qquad (9)$$

According to the analysis in [47], the effect of the low frequency dominate mode should also be considered at higher frequencies. Thus, the surface admittance of the first GPP is considered at the second band as well.

$$Y_{in}\mid_{\omega=\omega_2} = (R_1 + jL_1\omega_2 + \frac{1}{jC_1\omega_2})^{-1} + (R_3 + jL_3\omega_2 + \frac{1}{jC_3\omega_2})^{-1} - jY_s\cot(\beta_s h)\mid_{\omega=\omega_2}, \qquad (10)$$

To design a dual-band absorber, the following conditions should be considered at the second band:

$$\beta_s h\mid_{\omega=\omega_2} = \pi/2, \qquad (11.a)$$

$$L_3 C_3 = \frac{1}{\omega_2^2}, \qquad (11.b)$$

$$R_3 = \eta_0/\alpha. \qquad (11.c)$$

The $\alpha$ parameter will be defined later in this section. At sufficiently low frequencies, where the inter-band term of conductivity is dominant and for $E_F \gg k_B T$, $\sigma_g$ will be of the Drude form:

$$\sigma_g = \frac{e^2 E_F \tau}{\pi \hbar^2} \frac{1}{1 + j\omega\tau}. \qquad (12)$$

The values of $R$, $L$, and $C$ are obtained as:

$$R_n = \frac{L}{S_n^2} \frac{\pi \hbar^2}{e^2 E_F \tau}, \qquad (13.a)$$

$$L_n = \tau R_n, \qquad (13.b)$$

$$C_n = \frac{S_n^2}{L} \frac{2\varepsilon_{\text{eff}}}{q_n}. \qquad (13.c)$$

Therefore, (11) results in:

$$h = \frac{c}{4n_s f_2}, \qquad (14.a)$$

$$w = \frac{e^2 E_f r_3}{2\pi \hbar^2 \varepsilon_{\text{eff}} \omega_2^2}, \qquad (14.b)$$

$$\tau E_f = \alpha \frac{L}{w} \frac{\pi \hbar^2}{\varsigma_3 e^2 \eta_0}, \qquad (14.c)$$

where, $r_n = q_n w$ depends on the filling factor ($w/L$) and $\varsigma_n = S_n^2/w$. In addition, $\varsigma_1 = 0.88$ and $\varsigma_3 = 0.0573$ are obtained using (3). Hence, based on (11), the input admittance at $\omega_2$ is obtained as:

$$Y_{in} \mid_{\omega=\omega_2} = (R_1 + jL_1\omega_2 + \frac{1}{jC_1\omega_2})^{-1} + \frac{1}{R_3}. \tag{15}$$

Using $L_1 C_1 = \frac{r_3}{r_1\omega_2^2}$ and $L_1 = \tau R_1$ in (15), we obtain:

$$Y_{in} \mid_{\omega=\omega_2} = \frac{1}{R_3} \left( 1 + \frac{R_3 \left(1 - j\tau\omega_2(1 - \frac{r_1}{r_3})\right)}{R_1\left(1 + \tau^2\omega_2^2(1 - \frac{r_1}{r_3})^2\right)} \right). \tag{16}$$

To achieve more than 95% absorption, $|S_{11}|^2 < 0.05$ should be satisfied thus,

$$|S_{11}| = \left| \frac{1 - Y_{in}\eta_0}{1 + Y_{in}\eta_0} \right| < 0.22. \tag{17}$$

Using $\frac{R_3}{R_1} = \frac{S_1^2}{S_3^2}$ and $R_3 = \eta_0/\alpha$, (17) results in:

$$\left| \frac{1 - \left( \frac{S_1^2\left(1 - j\tau\omega_2(1 - \frac{r_1}{r_3})\right)}{S_3^2\left(1 + \tau^2\omega_2^2(1 - \frac{r_1}{r_3})^2\right)} + 1 \right)\alpha}{1 + \left( \frac{S_1^2\left(1 - j\tau\omega_2(1 - \frac{r_1}{r_3})\right)}{S_3^2\left(1 + \tau^2\omega_2^2(1 - \frac{r_1}{r_3})^2\right)} + 1 \right)\alpha} \right| < 0.22. \tag{18}$$

Therefore, $\alpha$ should be chosen in a way that the absorption is larger than 95% at the second band. Now, we impose the conditions to achieve perfect absorption at the first absorption frequency:

$$\text{Im}(Y_{in})\big|_{\omega=\omega_1} = 0, \tag{19.a}$$

$$\text{Re}(Y_{in})\big|_{\omega=\omega_1} = 1/\eta_0. \tag{19.b}$$

Therefore,

$$\frac{L_1\omega_1(1 - \frac{1}{L_1 C_1 \omega_1^2})}{R_1^2 + L_1^2\omega_1^2(1 - \frac{1}{L_1 C_1 \omega_1^2})^2} + Y_s \cot(\beta_s h)\big|_{\omega=\omega_1} = 0, \tag{20.a}$$

$$\frac{R_1}{R_1^2 + L_1^2\omega_1^2(1 - \frac{1}{L_1 C_1 \omega_1^2})^2} = 1/\eta_0. \tag{20.b}$$

Using $L_1 C_1 = \frac{r_3}{r_1\omega_2^2}$, $R_3 = \eta_0/\alpha$ and $L_1 = \tau R_1$ in (20), we obtain:

$$\tau\omega_1(1 - \frac{r_1\omega_2^2}{r_3\omega_1^2}) = -n_s \cot(\beta_s h)\big|_{\omega=\omega_1}, \tag{21.a}$$

$$\left(\tau\omega_1(1 - \frac{r_1\omega_2^2}{r_3\omega_1^2})\right)^2 = \frac{\alpha R_3}{R_1} - 1, \tag{21.b}$$

which results in

$$\left(n_s \cot(\beta_s h)\big|_{\omega=\omega_1}\right)^2 = \frac{\alpha S_1^2}{S_3^2} - 1. \tag{22}$$

So, the ratio between the first and second resonance frequencies is a constant defined as:

$$\frac{\omega_1}{\omega_2} = \frac{2}{\pi} \cot^{-1}\left( \frac{1}{n_s} \sqrt{\frac{\alpha S_1^2}{S_3^2} - 1} \right). \tag{23}$$

Based on the above analysis, the design procedure of the dual-band absorber in figure 1 is summarized as follows: First, we consider a specified value for the second resonance frequency ($\omega_2$). Then, $\alpha$ is determined using (18). Note that $r_1$ and $r_3$ should be specified in (18). Therefore, a practical value should be considered for the filling factor ($w/L$). The relaxation time of graphene and the first resonance frequency are obtained using (21) and (23) respectively. After calculating the graphene relaxation time, the Fermi level is obtained using (14.b). Finally, the width of the graphene ribbons and the array period are calculated using (14.b) and the specified value of the filling factor. It is worth mentioning that the Fermi level should be within $0 - 1$ eV. Furthermore, the relation between the relaxation time and the Fermi level is defined as:

$$\tau = \frac{E_f \mu}{e \, v_f^2}, \tag{24}$$

where $\mu$ is the electron mobility ranging from 0.03 $m^2$/Vs to 20 $m^2$/Vs depending on the fabrication process [50-53] and $v_f = 10^6$ m/s is the Fermi velocity [37]. Hence, the electron mobility range restricts the Fermi level selection.

The above design procedure is used to design three dual-band absorbers with the second resonance frequencies of $f_2 = 3$ THz, $f_2 = 4$ THz, and $f_2 = 5$ THz. The dielectric spacer is considered $Al_2O_3$ with $n_s = 3.13$. Table. I lists the designed parameters for proposed absorbers, when the filling factor is $w/L = 0.6$ and $\alpha = 1$. For this filling factor, $r_1 = 0.62\pi$ and $r_3 = 2.735\pi$ according to the data in [47]. Figure 3 shows the simulated absorption responses for the designed absorbers. The comparisons between the results obtained using the analytical circuit model and the ones from the full-wave simulations in HFSS show a good agreement verifying the developed design procedure. In the full-wave simulations, graphene is modelled as a $\Delta = 1$ nm thick layer with a relative permittivity of $\varepsilon = \varepsilon_0 - j\sigma_g/(\Delta\omega)$. The designed absorbers show

high-$Q$ absorption at both of the absorption bands. In Table. I, the quality factor ($Q$) is defined as:

$$Q = \frac{f_0}{\Delta f_{50\%}} \qquad (25)$$

where $\Delta f_{50\%}$ is the 50% absorption bandwidth at each band and $f_0$ is the central frequency of each band. Furthermore, the absorption is smaller than 0.02 in the frequency range between the two absorption bands showing a good out-of-band absorption characteristic.

Figures 4(a) and (b) show the simulated loss distributions on a single ribbon at the first and second resonance frequencies, respectively. It is clear that the two first GPPs are excited at the two resonance frequencies. In the first band, the loss distribution is focused at the center of the ribbon and decreases at the sides of the ribbons. The GPP at the second resonance frequency corresponds to the third eigenfunction in the array of the ribbons. As shown in figure 4(b), for the second band, the losses are concentrated at three main areas (the center and two sides of the center). The loss distributions are uniform along $y$-direction.

Table. I: The parameters of the dual-band proposed absorbers

| $f_2$ (THz) | $f_1$ (THz) | $\tau$ (ps) | $E_f$ (eV) | $h$ ($\mu$m) | $w$ ($\mu$m) | $L$ ($\mu$m) | Lower / upper $Q$-factor |
|---|---|---|---|---|---|---|---|
| 3 | 1.32 | 2.64 | 0.25 | 8 | 7.4 | 12.3 | 12.8 / 22.5 |
| 4 | 1.76 | 2 | 0.33 | 6 | 5.5 | 9.2 | 13.1 / 21.3 |
| 5 | 2.2 | 1.6 | 0.41 | 4.8 | 4.4 | 7.4 | 13.1 / 21.7 |

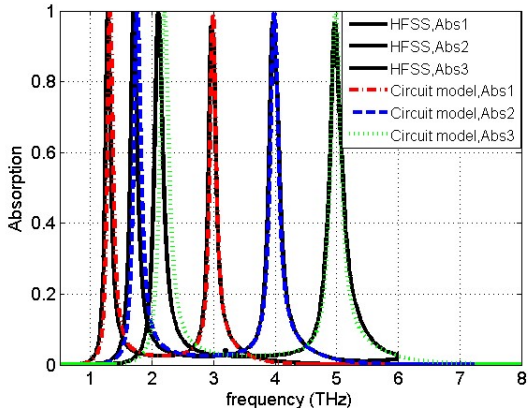

**Figure 3.** Absorption responses of the absorbers based on graphene ribbon array with the parameters presented in Table. I.

The influence of the Fermi level changes on the absorption response is investigated in figure 5. Based on the results, the absorption increases first and then decreases by increasing the Fermi level. However, the changes are more severe for the absorption peak of the first band. Furthermore, the second resonance frequency varies considerably by alerting the Fermi level, whereas the variations are smaller for the first resonance frequency.

The proposed absorber is only applicable for the TM polarization since the ribbons array is inductive for the TE polarization. Additionally, the grounded dielectric spacer is inductive within the whole frequency range. Therefore, the input impedance is inductive in the TE mode and resonance conditions are not met. In order to realize an insensitive absorber to the polarization of the incident wave, a two-dimensional patterned array with a symmetric geometry along $x$- and $y$-directions should be considered. Such a structure is designed in the next section using an array of graphene disks.

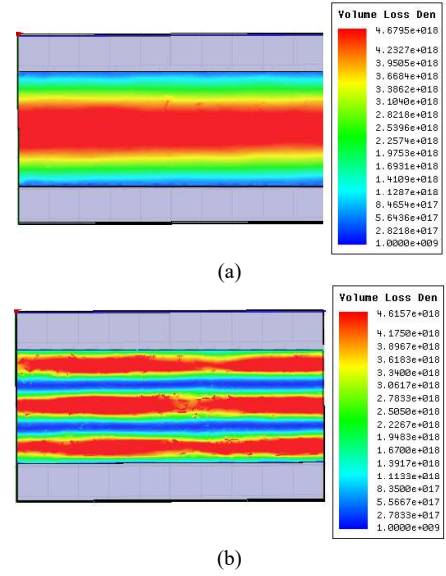

**Figure 4.** The loss density distributions on the graphene ribbon at two resonance frequencies. (a) The first and (b) the second resonance band.

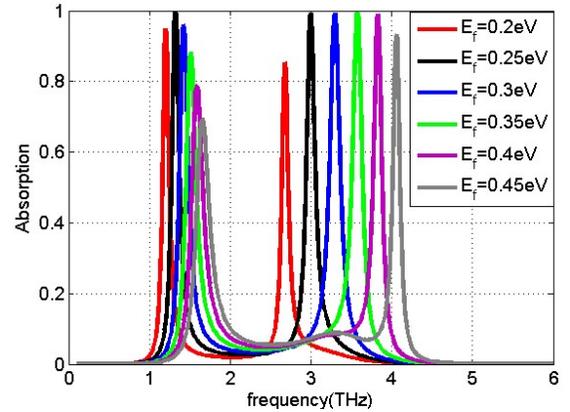

**Figure 5.** Influence of the Fermi level changes on the absorption spectra of the absorber 1 presented in Table. I.

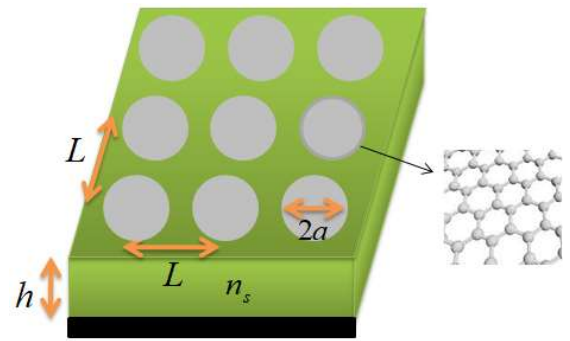

**Figure 6.** An absorber composed of an array of graphene disks with radius $a$ and period $L$ printed on a ground plane backed dielectric spacer. The graphene layer is shown with gray, whereas the dielectric spacer is shown with green and the ground layer is shown in black color.

## 3. Dual-band absorber based on an array of graphene disks

Consider an array of graphene disks printed on a ground plane backed dielectric spacer shown in figure 6. The periodicity of the structure along $x$- and $y$-directions is $L$ and the radius of the graphene disk is $a$. An equivalent circuit model of the absorber is demonstrated in figure 2 for a normally incident plane wave. The equivalent surface admittance of the disks array is analytically calculated as [48]:

$$Y_{sur} = \frac{\pi^2}{L^2} \sum_{n=1}^{\infty} \frac{S_n^2}{K_n^2} \left( \sigma_g^{-1} + \frac{q_{1n}}{j\omega\varepsilon_{eff}} \right)^{-1}, \tag{26}$$

where

$$S_n = \int_0^a \left[ \frac{df_{1n}(\rho)}{d\rho} \rho + f_{1n}(\rho) \right] d\rho, \tag{27}$$

with $f_{1n}(p)$ given in [48] and

$$K_n = \int_S \xi_{1n} \cdot \xi_{1n}^* \, dS. \tag{28}$$

In (27), $\xi_{1n} = \nabla \left[ f_{1n}(\rho) \cos \varphi \right]$ and $S$ is the surface area of a disk. The $q_{1n}$ coefficient is an eigenvalue of the equation governing the current distribution on the disks for various $2a/L$ [48]. The two first eigenfunctions are calculated as:

$$f_{11}(\rho) = [J_1(\xi_{11}\rho/a) - 0.06 J_1(\xi_{12}\rho/a) + \\ 0.022 J_1(\xi_{13}\rho/a))], \tag{29.a}$$

$$f_{12}(\rho) = [-0.25 J_1(\xi_{11}\rho/a) - 0.965 J_1(\xi_{12}\rho/a) \\ + 0.085 J_1(\xi_{13}\rho/a))], \tag{29.b}$$

where the corresponding eigenvalues are $q_{11} = 0.417/a$ and $q_{12} = 2.426/a$ for $2a/L = 0.9$, respectively [48]. As in figure 2(b), the surface admittance presented in (26) can be considered as parallel branches composed of the $RLC$ series tanks corresponding to the GPPs on the disks. The lumped element values are calculated as:

$$R_n = \frac{L^2}{\pi^2} \frac{K_n}{S_n^2} \mathrm{Re}\{\sigma_g^{-1}\}, \tag{30.a}$$

$$L_n = \frac{L^2}{\pi^2} \frac{K_n}{S_n^2} \frac{\mathrm{Im}\{\sigma_g^{-1}\}}{\omega}, \tag{30.b}$$

$$C_n = \frac{\pi^2}{L^2} \frac{S_n^2}{K_n} \frac{\varepsilon_{eff}}{q_{1n}}. \tag{30.c}$$

Following the design process in the previous section, the ratio between the two resonance frequencies is defined as:

$$\frac{\omega_1}{\omega_2} = \frac{2}{\pi} \cot^{-1} \left( \frac{1}{n_s} \sqrt{\frac{\alpha S_1^2 K_2}{S_2^2 K_1} - 1} \right), \tag{31}$$

where $S_1 = 0.612a$, $S_2 = 0.2302a$, $K_1 = 1.2891$ and $K_2 = 4.96$ obtained using (27) and (28). The relaxation time of graphene is calculated as:

$$\tau = \frac{-n_s \cot(\beta_s h)|_{\omega=\omega_1}}{\omega_1 (1 - \frac{r_{11}}{r_{12}} \frac{\omega_2^2}{\omega_1^2})}, \tag{32}$$

where $r_{1n} = q_{1n}a$. For a specified value of $\alpha$, the center frequency of the first band and the relaxation time of graphene are obtained using (31) and (32). However, the following condition should be checked to achieve more than 95% absorption at the second band.

$$\left| 1 - \frac{\left( \frac{K_2 S_1^2 \left(1 - j\tau\omega_2(1 - \frac{r_{11}}{r_{12}})\right)}{K_1 S_2^2 \left(1 + \tau^2\omega_2^2(1 - \frac{r_{11}}{r_{12}})^2\right)} + 1 \right) \alpha}{1 + \frac{\left( \frac{K_2 S_1^2 \left(1 - j\tau\omega_2(1 - \frac{r_{11}}{r_{12}})\right)}{K_1 S_2^2 \left(1 + \tau^2\omega_2^2(1 - \frac{r_{11}}{r_{12}})^2\right)} + 1 \right) \alpha} \right| < 0.22. \tag{33}$$

Finally, the geometrical parameters of the structure are extracted as:

$$a = \frac{e^2 E_f r_{12}}{\pi \hbar^2 \varepsilon_{eff} \omega_2^2}, \tag{34.a}$$

$$\tau E_f = \alpha \left( \frac{L}{a} \right)^2 \frac{K_2 \pi \hbar^2}{\pi^2 \zeta_2^2 e^2 \eta_0}, \tag{34.b}$$

where $\zeta_2 = S_2/a$.

Here, a dual-band absorber is designed based on an array of disks. As an example, suppose that the center frequency of the second band is $f_2 = 5$ THz. The geometrical parameters of the absorber are obtained as: $a = 2.3$ $\mu m$, $L = 5.1$ $\mu m$, and $h = 4.8$ $\mu m$. The properties of graphene are calculated as $\tau = 2.2$ ps and $E_f = 0.38$ eV. In the absorber design, $\alpha = 0.8$ is considered to do the calculations satisfying (33). The analytical results are verified by comparisons with the results obtained using the full-wave simulations in HFSS. The absorption response of the structure is plotted in figure 7 as a function of frequency. The absorption levels are 100% and 95% at the first and second bands, respectively showing a good agreement with the designed values. The simulated loss distributions on the disks are presented in figure 8 at both of the resonance frequencies for a normal incident angle. For the first resonance frequency, the losses are concentrated in the central area of the disc, whereas there are additional two side areas with high loss distributions at the second resonance frequency. As explained in [39], the Fermi level of graphene can be regulated in a specific range by applying a bias voltage. Figure 9 shows the absorption spectra for a range of fermi levels around the designed 0.38 eV (0.4 eV - 0.6 eV). As seen, the first resonance frequency is slightly shifted to higher frequencies by increasing the Fermi level with considerable variations in the absorption peak. However, the variation of the second resonant frequency is large with slight changes in the peak absorption.

Table. II. Comparison results of presented absorber with other similar dual-band graphene absorbers at terahertz frequencies.

| Ref. | No. of Graphene layers /Configuration | Abs. bands (THz) | Lower / upper band Q factor | Inter band absorption level | Polarization Sensitive | Absorption >0.9 Angle Range |
|---|---|---|---|---|---|---|
| [27] | 2/ graphene disk array | 3 and 5 | 22 / 41 | Below 0.02 | No | 0° to 20° |
| [39] | 1/ Periodic array of single sizegraphene patches | 0.53 and 1.53 | 2.4 / 5.1 | 0.3 | No | 0° to 60° |
| [54] | 2/graphene sheet and ribbon array | 4.95 and 9.2 | 7.3 / 12.8 | 0.1 | Yes | N.A. |
| [55] | 1/Two different size garphene disks in a unit cell | 7.1 and 10.4 | 9.7 / 11.1 | 0.14 | No | 0° to 70° |
| [56] | 1/Horizontal and vertical elliptical graphene disks in a unit cell | 5 and 8.57 | 7.7 / 10.5 | 0.2 | No | 0° to 60° |
| This work | 1/ Periodic array of single sizegraphene disks | 1.9 and 5 | 16 / 40 | Below 0.02 | No | 0° to 70° |

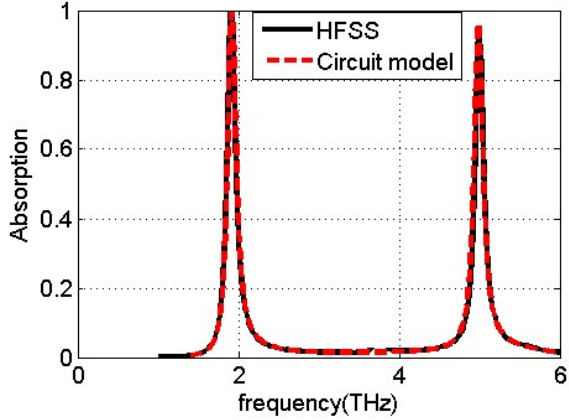

**Figure 7.** Absorption spectra of the absorber in figure 6 with parameters $a$ = 2.3 $\mu$m, $L$ = 5.1 $\mu$m, and $h$ = 4.8 $\mu$m. The properties of graphene are $\tau$ = 2.2 ps and $E_f$ = 0.38 eV.

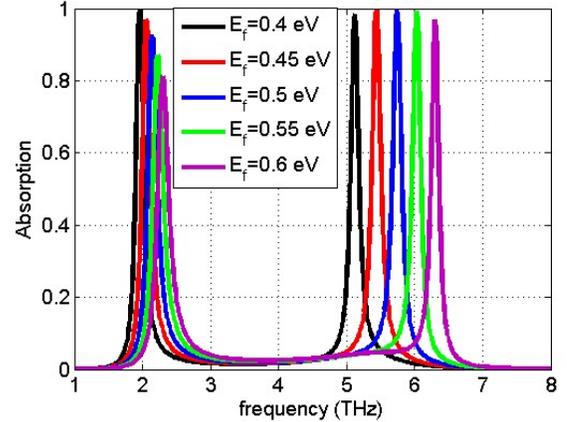

**Figure 9.** Absorption spectra of the proposed graphene-based absorber for different Fermi levels.

In order to investigate the influence of the incidence angle on the absorption properties of the proposed absorber, the absorption spectra is simulated for different incidence angles as plotted in figure 10. For the incidence angles below 70°, the absorption level remains higher than 90% in both of the TE and TM polarizations.

A comparison between the designed absorber and other similar dual-band graphene-based terahertz absorbers is presented in Table. II. The absorbers in [27] and [54] are made of two stacked graphene layers requiring a complicated multi-layer fabrication. On the other hand, the unit cells of the absorbers in [27] and [54-56] are made of complicated geometries that might be challenging to implement using graphene. The dual-band absorbers in [55] and [56] are based on combining two resonators with different geometries to generate dual resonance and absorption bands. However, the proposed absorber is composed of a simple graphene disk element in a unit cell. In addition, in comparison with the rest of the designs, the absorption bands show higher $Q$-factors in the proposed absorber. The designed absorber offers a highly stable frequency response up to 70° oblique incidence angle for both of the TE and TM polarizations. Only the absorbers in [55, 56] show a similar scan angle performance. Furthermore, the interband absorption level is smaller than 0.02 for the proposed absorber showing a superior performance with respect to the majority of the designs.

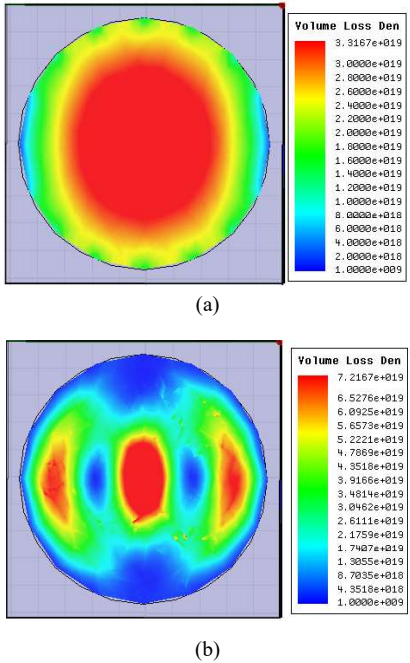

**Figure 8.** The loss density distributions on the graphene disks. (a) Distribution at the first resonance frequency and (b) at the second resonance frequency.

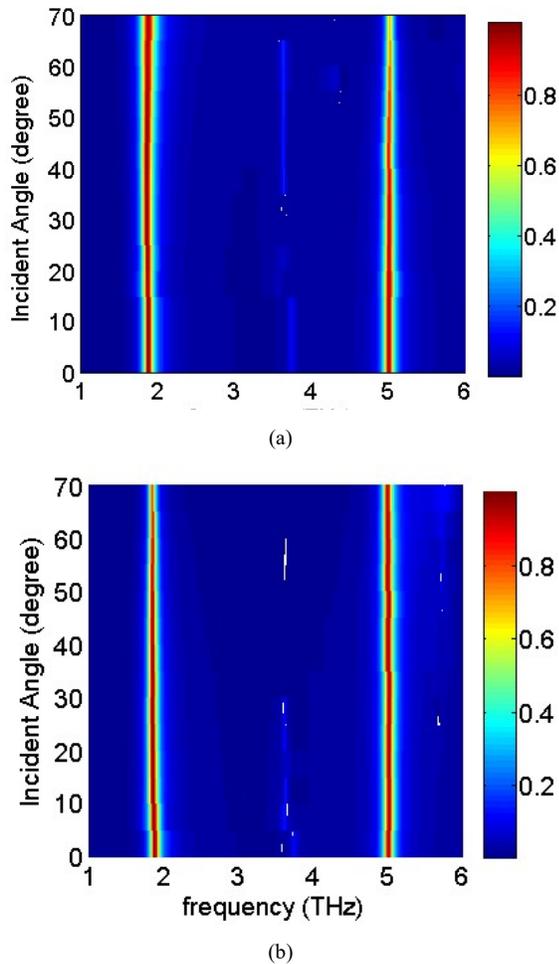

(a)

(b)

**Figure 9.** Absorption spectra of the disc array graphene-based absorber as function of incident angle and frequency for (a) TM and (b) TE polarizations.

## 4. Conclusion

Dual-band absorbers based on graphene ribbons and disks printed on a grounded dielectric slab have been investigated for low terahertz regimes. The design is based on selectively exciting the two first GPPs to achieve dual absorption bands. The circuit models including two parallel branches of the *RLC* series circuits corresponding to the GPPs are used for analytical design of the absorbers. Closed-form relations are obtained for proper design of the geometrical parameters and graphene properties. The simulated loss distributions at the two resonance frequencies confirmed the excitation of a specific GPP at each resonanceband. The simulations verify the controllability of the absorber resonance frequencies by controlling the geometrical parameters and the Fermi level. The absorber designed using the graphene disk array shows a polarization insensitive response with a high absorption stability over a wide range of scan angles up to 70°. These are achieved together with high-Q absorption at both bands.